\documentclass[aps,pra,showpacs,twocolumn]{revtex4}%
\usepackage{amsmath}
\usepackage{amsfonts}
\usepackage{amssymb}
\usepackage{graphicx}
\usepackage[colorlinks,linkcolor=blue,anchorcolor=blue,citecolor=blue,urlcolor=black]%
{hyperref}
\usepackage{mathrsfs}
\usepackage{dcolumn}
\usepackage{bm}
\usepackage{epsfig}%
\providecommand{\U}[1]{\protect\rule{.1in}{.1in}}

\begin{document}
\title{Optimal limits of cavity optomechanical cooling in the strong coupling regime}
\author{Yong-Chun Liu$^{1,2}$}
\author{Yu-Feng Shen$^{1}$}
\author{Qihuang Gong$^{1,2}$}
\author{Yun-Feng Xiao$^{1,2}$}
\email{yfxiao@pku.edu.cn}
\altaffiliation{URL: www.phy.pku.edu.cn/$\sim$yfxiao/index.html}

\affiliation{$^{1}$State Key Laboratory for Mesoscopic Physics and School of Physics,
Peking University, Beijing 100871, P. R. China}
\affiliation{$^{2}$Collaborative Innovation Center of Quantum Matter, Beijing 100871,
People's Republic of China}
\date{\today}

\begin{abstract}
Laser cooling of mesoscopic mechanical resonators is of great interest for
both fundamental studies and practical applications. We provide a general
framework to describe the cavity-assisted backaction cooling in the strong
coupling regime. By studying the cooling dynamics, we find that the temporal
evolution of mean phonon number oscillates as a function of the optomechanical
coupling strength depending on frequency mixing. The further analytical result
reveals that the optimal cooling limit is obtained when the system eigenmodes
satisfy the frequency matching condition. The reduced instantaneous-state
cooling limits with dynamic dissipative cooling approach are also presented.
Our study provides a guideline for optimizing the backaction cooling of
mesoscopic mechanical resonators in the strong coupling regime.

\end{abstract}

\maketitle

\section{Introduction}

Cavity optomechanics \cite{Rev}, which explores the interaction between light
and mechanical motion, provides a unique platform for various applications,
such as the fundamental test of quantum theory \cite{super}, quantum
information processing \cite{QIP} and high-precision measurements \cite{MM}.
Recent theoretical and experimental efforts have demonstrated optomechanically
induced transparency \cite{OMIT}, optomechanical storage \cite{StorPRL11},
normal mode splitting \cite{SCPRL08,SCNat09,SCNat11}, quantum-coherent
coupling between optical modes and mechanical modes \cite{SCNat12,SCNat13},
state transfer at different optical (electromagnetic) wavelengths \cite{ST},
quantum entanglement \cite{Ent}, squeezing \cite{Sqz} and nonlinear quantum
optomechanics \cite{NL}. For most applications, it is a prerequisite to cool
the mechanical resonators close to the quantum ground state so as to suppress
the thermal noise. In the past few years, numerous experiments have
demonstrated cooling of mechanical resonators by employing pure cryogenic
cooling \cite{GSNat10}, feedback cooling (active cooling, or cold-damping)
\cite{FBCoo} and backaction cooling (passive cooling, or self-cooling)
\cite{CooNat06,CooNat06-2,CooPRL06,CooNatPhys08,CooNatPhys09-1,CooNatPhys09-2,CooNatPhys09-3,CooNat10,GSNat11,GSNat11-2,CooPRA11}%
. The backaction cooling is proved to be efficient
\cite{PRL07-1,PRL07-2,PRA08}, especially in the resolved sideband limit, where
the mechanical resonance frequency is greater than the decay rate of the
optical cavity. Recently many efforts have been taken to extend or improve the
backaction cooling, for example, cooling with dissipative coupling \cite{DC},
cooling with quadratic coupling \cite{quadCooPRA10}, cooling with hybrid
systems \cite{Atom}, cooling in the single photon strong coupling regime
\cite{SSCCooPRA12} and pulsed laser cooling \cite{Pul}.

Besides ground state cooling, another crucial condition for quantum operation
is strong coupling \cite{SCNat12,SCNat13}, where the light-enhanced
optomechanical coupling strength exceeds the cavity decay rate. In such a
regime, the optical and mechanical modes hybridize, leading to normal mode
splitting \cite{SCPRL08,SCNat09,SCNat11}. In the time domain, the energy
exchange between the optical and mechanical modes is reversible, which allows
state swapping. However, it also brings about swap heating, which leads to the
saturation of cooling rates. To overcome this problem, we have recently
\cite{ycliuDC13} proposed a dynamic dissipative cooling approach, which
significantly accelerates the cooling process and reduces the cooling limits.
In this paper we extend our previous results and present a detailed
exploration of the cooling dynamics and cooling limits in the strong coupling
regime. We systematically investigate the cooling dynamics under the
rotating-wave approximation (RWA) and without RWA. Rabi-like oscillations and
frequency mixing phenomenon in the time evolution of mean phonon number are
studied. To obtain the lowest cooling limits, we analytically derive the
frequency matching condition and a small cavity decay rate is preferred.

The rest of this paper is organized as follows. In Sec. II we use the
linearized quantum Langevin equations and quantum master equation to describe
the system. In Sec. III we discuss the simplified model under RWA. In Sec. IV
we study the full cooling dynamics without the RWA, where the zero-temperature
and finite-temperature cases are considered in succession. In Sec. V we
discuss the reduced instantaneous-state cooling limits with dynamic
dissipative cooling approach. A summary is presented in Sec. VI.

\section{Theoretical Model}

\begin{figure}[t]
\centerline{\includegraphics[width=\columnwidth]{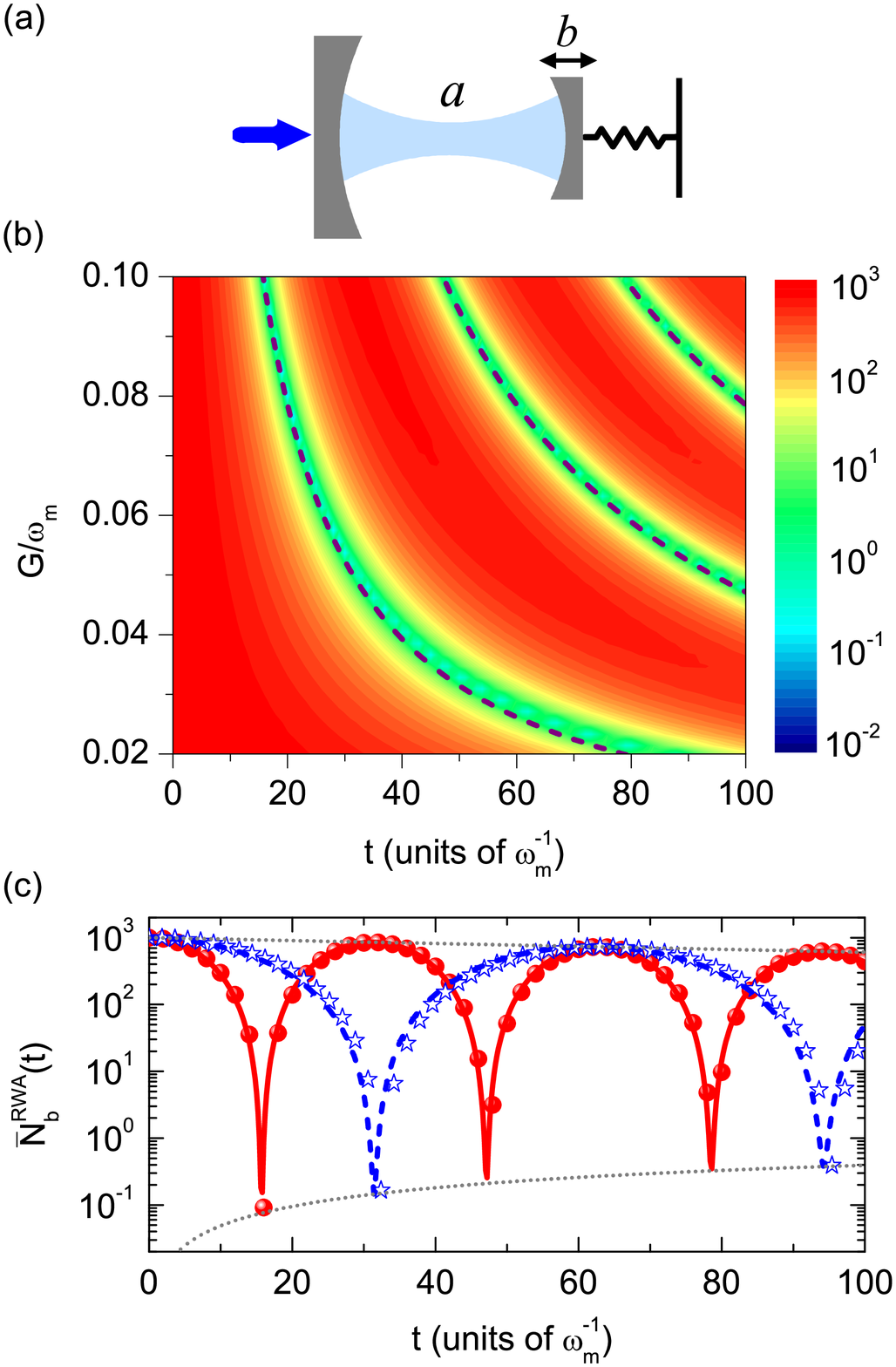}}
\caption{(color online) (a) Sketch of the optomechanical system. (b) Time
evolution of the mean phonon number $\bar{N}_{b}^{\mathrm{RWA}}(t)$ (under
RWA) as functions of $t$ and $G$ for $\Delta^{\prime}=-{\omega_{\mathrm{m}}}$,
${\kappa/\omega_{\mathrm{m}}=0.01}$, ${\gamma/\omega_{\mathrm{m}}=10}^{-5}$
and $n_{\mathrm{th}}=10^{3}$. The purple dashed curves correspond to the cases
of $t=\pi/(2\left\vert {G}\right\vert )$, $3\pi/(2\left\vert {G}\right\vert )$
and $5\pi/(2\left\vert {G}\right\vert )$. The color bar is in log scale. (c)
$\bar{N}_{b}^{\mathrm{RWA}}(t)$ as a function of $t$ for $G{/\omega
_{\mathrm{m}}}=0.1$ (red circles and red solid curve) and $0.05$ (blue stars
and blue dashed curve). The circles and stars are numerical results obtained
from Eq. (\ref{DRWA}) and the curves are analytical results calculated from
Eq. (\ref{NbRWA}). The gray dotted curves denote the envelopes given by
$n_{\mathrm{th}}{e}^{-\frac{{\kappa+\gamma}}{2}t}$ (upper curve) and
$n_{\mathrm{th}}{(1-e}^{-\frac{{\kappa+\gamma}}{2}t}){\gamma}/({\kappa+\gamma
})$ (lower curve).}%
\label{Fig1}%
\end{figure}

We consider a typical optomechanical system involving an optical mode and a
mechanical mode, with a coherent laser driving the optical cavity [Fig.
\ref{Fig1}(a)]. The system Hamiltonian reads $H={{\omega}_{\mathrm{c}}a^{\dag
}a}+{\omega_{\mathrm{m}}b^{\dag}b+ga^{\dag}a{(b+b^{\dag})+({\Omega e}}%
}^{-i\omega t}{a^{\dag}+\Omega}^{\ast}{{{e}}}^{i\omega t}{a).}$Here the first
(second) term describes the energy of the optical (mechanical) mode, with the
angular resonance frequency ${\omega}_{\mathrm{c}}$ (${\omega_{\mathrm{m}}}$),
the annihilation operator ${a}$ (${b}$) and the creation operator ${a^{\dag}}$
(${b^{\dag}}$). The third term represents the optomechanical interaction
\cite{LawPRA95}, with the single-photon optomechanical coupling rate $g$. The
last term describes the driving of the input laser, where ${{\Omega=}}%
\sqrt{\kappa_{\mathrm{ex}}P/(\hbar\omega)}e^{i\phi}$ denotes the driving
strength with the input laser power $P$, initial phase $\phi$, frequency
$\omega$ and the input-cavity coupling rate $\kappa_{\mathrm{ex}}$. In the
interaction picture, the quantum Langevin equations are given by
\begin{subequations}
\begin{align}
\dot{a}  &  =\left(  i{\Delta}-\frac{\kappa}{2}\right)  a-i{{g}a{(b+b^{\dag}%
)}}\nonumber\\
&  {-i\Omega-}\sqrt{\kappa_{\mathrm{ex}}}a_{\mathrm{in,ex}}-\sqrt
{\kappa_{\mathrm{0}}}a_{\mathrm{in,0}}{,}\label{a}\\
\dot{b}  &  =\left(  -i{\omega_{\mathrm{m}}}-\frac{\gamma}{2}\right)
b-i{{g}a^{\dag}a-}\sqrt{\gamma}b_{\mathrm{in}}, \label{b}%
\end{align}
where ${\Delta}=\omega-{{\omega}_{\mathrm{c}}}$ denotes the input-cavity
detuning; $\kappa_{\mathrm{0}}$ stands for the intrinsic cavity dissipation
rate; $\kappa=\kappa_{\mathrm{0}}+\kappa_{\mathrm{ex}}$ represents the total
cavity dissipation rate; $\gamma$ is the dissipation rate of the mechanical
mode; $a_{\mathrm{in,0}}$, $a_{\mathrm{in,ex}}$ and $b_{\mathrm{in}}$ are the
noise operators associated with the intrinsic cavity dissipation
$\kappa_{\mathrm{0}}$, external cavity dissipation $\kappa_{\mathrm{ex}}$ and
mechanical dissipation $\gamma$, which have zero mean values and obey the
correlation functions $\langle a_{\mathrm{in,0}}(t)a_{\mathrm{in,0}}^{\dag
}(t^{\prime})\rangle=\langle a_{\mathrm{in,ex}}(t)a_{\mathrm{in,ex}}^{\dag
}(t^{\prime})\rangle=\delta(t-t^{\prime})$, $\langle a_{\mathrm{in,0}}^{\dag
}(t)a_{\mathrm{in,0}}(t^{\prime})\rangle=\langle a_{\mathrm{in,ex}}^{\dag
}(t)a_{\mathrm{in,ex}}(t^{\prime})\rangle=0$, $\langle b_{\mathrm{in}%
}(t)b_{\mathrm{in}}^{\dag}(t^{\prime})\rangle=(n_{\mathrm{th}}+1)\delta
(t-t^{\prime})$ and $\langle b_{\mathrm{in}}^{\dag}(t)b_{\mathrm{in}%
}(t^{\prime})\rangle=n_{\mathrm{th}}\delta(t-t^{\prime})$. Here
$n_{\mathrm{th}}=[\exp(\frac{\hbar{\omega_{\mathrm{m}}}}{k_{\mathrm{B}}%
T})-1]^{-1}$ describes the equilibrium mean thermal phonon number, where $T$
is the temperature of the reservoir and $k_{\mathrm{B}}$ is Boltzmann
constant. For strong coherent laser input, both the optical and mechanical
modes reach a new steady state, and thus we can rewrite the operators as
${a\rightarrow\alpha+a}_{1}$ and $b\rightarrow\beta+b_{1}$. Here ${\alpha}$
and $\beta$ represent the $c$-number steady state values of the optical and
mechanical modes, while ${a}_{1}$ and $b_{1}$ are the corresponding
fluctuation operators. From Eqs. (\ref{a}) and (\ref{b}), the quantum Langevin
equations for the quantum fluctuations are given by
\end{subequations}
\begin{subequations}
\begin{align}
\dot{a}_{1}  &  =\left(  i{\Delta}^{\prime}-\frac{\kappa}{2}\right)
a_{1}-{{{i{{g}}\alpha{(b_{1}}}+{{b_{1}^{\dag})}}}}\nonumber\\
&  -ig{{{a}}}_{1}{{{{(b}}}}_{1}+{{{{b_{1}^{\dag})}}-}\sqrt{\kappa
_{\mathrm{ex}}}a_{\mathrm{in,ex}}-\sqrt{\kappa_{\mathrm{0}}}a_{\mathrm{in,0}%
},}\label{a1}\\
\dot{b}_{1}  &  =\left(  -i{\omega_{\mathrm{m}}}-\frac{\gamma}{2}\right)
b_{1}-ig\left(  {{{\alpha}}}^{\ast}{a_{1}+{{\alpha}}a_{1}^{\dag}}\right)
\nonumber\\
&  -i{{g}a_{1}^{\dag}a}_{1}-\sqrt{\gamma}b_{\mathrm{in}}, \label{b1}%
\end{align}
where ${\Delta}^{\prime}=\Delta-{{g({\beta}}}+{{{\beta}}}^{\ast}{{)}}$
represents the optomechanical-coupling modified detuning. For strong driving
$\left\vert {\alpha}\right\vert \gg1$, the nonlinear terms in Eqs. (\ref{a1})
and (\ref{b1}) can be neglected, yielding linearized quantum Langevin
equations. The corresponding quadratic Hamiltonian is given by
\end{subequations}
\begin{equation}
H_{L}=-\Delta^{\prime}{a_{1}^{\dag}a}_{1}+{\omega_{\mathrm{m}}{b_{1}^{\dag}%
b}_{1}+(Ga_{1}^{\dag}+G^{\ast}a_{1})(b_{1}+b_{1}^{\dag}}), \label{HL}%
\end{equation}
where $G={\alpha g}$ describes the light-enhanced optomechanical coupling
strength. In this Hamiltonian, the rotating-wave terms ${Ga_{1}^{\dag}%
b_{1}+G^{\ast}a_{1}b_{1}^{\dag}}$ correspond to the beam-splitter interaction,
while the counter-rotating-wave terms ${Ga_{1}^{\dag}b_{1}^{\dag}+G^{\ast
}a_{1}b_{1}}$ describe the two-mode squeezing interaction.

Starting from the Hamiltonian Eq. (\ref{HL}) in the linear regime, the time
evolution of the system density matrix $\rho$ is described by the quantum
master equation
\begin{align}
\dot{\rho}  &  =i[\rho,H_{L}]+\frac{\kappa}{2}\left(  2{a}_{1}\rho{a_{1}%
^{\dag}}-{{a_{1}^{\dag}{a}_{1}\rho-\rho a_{1}^{\dag}{a}_{1}}}\right)
\nonumber\\
&  +\frac{\gamma}{2}(n_{\mathrm{th}}+1)\left(  2{b}_{1}\rho{b_{1}^{\dag
}-{b_{1}^{\dag}b_{1}\rho-\rho b_{1}^{\dag}b_{1}}}\right) \nonumber\\
&  +\frac{\gamma}{2}n_{\mathrm{th}}\left(  2{b_{1}^{\dag}}\rho{b}_{1}%
-{{b_{1}{b_{1}^{\dag}}\rho-\rho b_{1}b_{1}^{\dag}}}\right)  . \label{Master}%
\end{align}
We focus on the time evolution of the mean phonon number $\bar{N}_{b}%
=\langle{b_{1}^{\dag}b}_{1}\rangle=\mathrm{Tr}(\rho{b_{1}^{\dag}b}_{1})$.
Using Eq. (\ref{Master}), $\bar{N}_{b}$ is determined by a linear system of
ordinary differential equations involving all the second-order moments
$\mathbf{V=(}\bar{N}_{a}$, $\bar{N}_{b}$, $\langle{a{_{1}^{\dag}{{{b_{1}}}}}%
}\rangle$, $\langle{a{_{1}{{{b_{1}^{\dag}}}}}}\rangle$, $\langle
{a{_{1}{{{b_{1}}}}}}\rangle$, $\langle{a{_{1}^{\dag}{{{b_{1}^{\dag}}}}}%
}\rangle$, $\langle{a{_{1}^{2}}}\rangle$, $\langle{a{_{1}^{\dag2}}}\rangle$,
$\langle{b{_{1}^{2}}}\rangle$, $\langle{b{_{1}^{\dag2}}}\rangle\mathbf{)}^{T}$
\cite{SCNJP08,ycliuDC13}, with the equations given by
\begin{equation}
\mathbf{\dot{V}=MV+N}, \label{D}%
\end{equation}
where $\bar{N}_{a}=\langle{a_{1}^{\dag}a}_{1}\rangle$, and the elements of the
matrices $\mathbf{M}$ and $\mathbf{N}$ are presented in the appendix.

\section{Cooling dynamics under the rotating-wave approximation}

We focus on the strong coupling regime, where the light-enhanced
optomechanical coupling strength $\left\vert {G}\right\vert $ is far greater
than the cavity decay rate $\kappa$. When $\left\vert {G}\right\vert
\ll{\omega_{\mathrm{m}}}$ is also satisfied, the rotating-wave approximation
can be made, so that we can concisely describe the main characteristic of the
cooling dynamics. Under the RWA, the counter-rotating-wave terms
${Ga_{1}^{\dag}b_{1}^{\dag}+G^{\ast}a_{1}b_{1}}$ are neglected, and Eq.
(\ref{D}) reduce to
\begin{subequations}
\label{DRWA}%
\begin{gather}
\frac{\partial\bar{N}_{a}}{\partial t}=-i|G|\bar{F}-{\kappa}\bar{N}_{a},\\
\frac{\partial\bar{N}_{b}}{\partial t}=i|G|\bar{F}-{\gamma}\bar{N}_{b}%
+{\gamma}n_{\mathrm{th}},\\
\frac{\partial\bar{F}}{\partial t}=-2i\left\vert G\right\vert \left(  \bar
{N}_{a}-\bar{N}_{b}\right)  -[i({\Delta}^{\prime}+{\omega_{\mathrm{m}}}%
)+\frac{{\kappa+\gamma}}{2}]F,
\end{gather}
where $\bar{F}=(G\langle{a{_{1}^{\dag}{{{b_{1}\rangle-}}}G}^{\ast}\langle
a_{1}{{{{b_{1}^{\dag}\rangle}}}}})/|G|$ represents the coherence between the
optical and mechanical modes. We consider the red sideband resonant case with
$\Delta^{\prime}=-{\omega_{\mathrm{m}}}$, \textit{i. e.}, the cooling process
is on resonance. The time evolution of the mean phonon number is given by
\end{subequations}
\begin{equation}
\bar{N}_{b}^{\mathrm{RWA}}(t)\simeq n_{\mathrm{th}}\frac{{\gamma+e}%
^{-\frac{{\kappa+\gamma}}{2}t}\left[  {\kappa\cos}^{2}(\left\vert
{G}\right\vert {t)-\gamma\sin}^{2}{(\left\vert {G}\right\vert t)}\right]
}{{\kappa+\gamma}}. \label{NbRWA}%
\end{equation}
It shows that the phonon number is proportional to the environmental thermal
phonon number $n_{\mathrm{th}}$, which reveals that the rotating-wave
optomechanical coupling only modifies the effective mechanical dissipation rate.

In Fig. \ref{Fig1}(b) we plot the exact numerical results of $\bar{N}%
_{b}^{\mathrm{RWA}}(t)$ as functions of $t$ and ${G}$ for ${\kappa
/\omega_{\mathrm{m}}=0.01}$, ${\gamma/\omega_{\mathrm{m}}=10}^{-5}$ and
$n_{\mathrm{th}}=10^{3}$, with Fig. \ref{Fig1}(c) plots both numerical and
analytical results of $\bar{N}_{b}^{\mathrm{RWA}}(t)$ for $G{/\omega
_{\mathrm{m}}}=0.1$ and $0.05$. It shows that the mean phonon number undergoes
Rabi-like oscillations with the period of $\pi/\left\vert {G}\right\vert $,
which implies the energy exchange between the optical mode and the mechanical
mode, together with the normal mode splitting with the frequency difference of
$2\left\vert {G}\right\vert $. The dissipation is characterized by the
envelopes, where the upper envelope is approximately described by
$n_{\mathrm{th}}{e}^{-\frac{{\kappa+\gamma}}{2}t}$, and the lower envelope
reads $n_{\mathrm{th}}{(1-e}^{-\frac{{\kappa+\gamma}}{2}t}){\gamma}%
/({\kappa+\gamma})$, as shown in Fig. \ref{Fig1}(c). From Figs. \ref{Fig1}(b)
and (c) we find small discrepancy for small $\left\vert {G}\right\vert $,
which is because the strong coupling condition is weakly satisfied for
$\left\vert {G}\right\vert \sim{\kappa}$. Taking the dissipations into
account, the normal mode splitting is given by $\sim2\sqrt{\left\vert
{G}\right\vert ^{2}-{\kappa}^{2}/16}$, which is slightly smaller than
$2\left\vert {G}\right\vert $. Thus the oscillation period is slightly larger
than $\pi/\left\vert {G}\right\vert $, which agrees with Figs. \ref{Fig1}(b)
and (c).

\begin{figure}[tb]
\centerline{\includegraphics[width=\columnwidth]{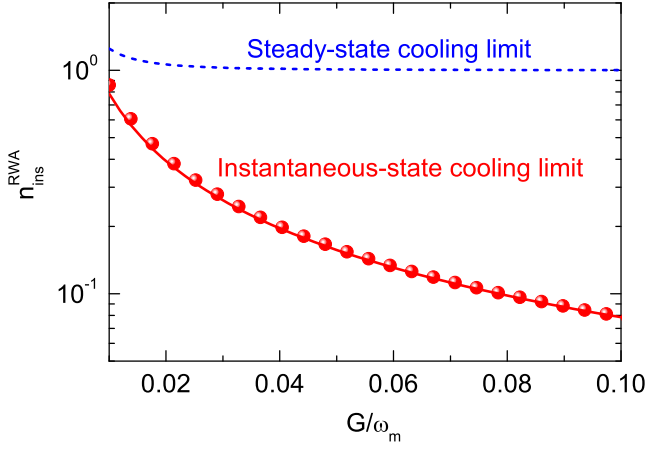}}
\caption{(color online) Instantaneous-state cooling limit $n_{\mathrm{ins}%
}^{\mathrm{RWA}}$ as a function of $G$. The red circles are numerical results
obtained from Eq. (\ref{DRWA}) and the red solid curve is the analytical
result calculated from Eq. (\ref{NinsRWA}). The steady-state cooling limit is
plotted for comparison (blue dashed curve). Other parameters are the same as
Fig. \ref{Fig1}(b).}%
\label{Fig2}%
\end{figure}

When the system reaches the steady state, the final phonon number is given by
$n_{\mathrm{th}}{\gamma}/({\kappa+\gamma})$, as inferred from Eq.
(\ref{NbRWA}). Nevertheless, the minimum phonon number is obtained near the
end of the first half Rabi oscillation cycle, ${t}\simeq\pi/(2G)$, with the
instantaneous-state cooling limit
\begin{equation}
n_{\mathrm{ins}}^{\mathrm{RWA}}\simeq\frac{\pi{\gamma}n_{\mathrm{th}}%
}{4\left\vert {G}\right\vert }. \label{NinsRWA}%
\end{equation}
It shows that the instantaneous-state cooling limit does not depend on the
cavity decay rate ${\kappa}$. This is a great improvement compared with the
steady-state cooling limit, which is constrained by the cavity decay rate. In
Fig. \ref{Fig2} we plot $n_{\mathrm{ins}}^{\mathrm{RWA}}$ as a function of
$G$. It reveals that a large coupling strength leads to a low cooling limit.
This is because the time to reach the minimum is in inverse proportional to
the coupling strength, and short evolution time suffers less dissipation, as
compared in Fig. \ref{Fig1}(c) for $G{/\omega_{\mathrm{m}}}=0.1$ and $0.05$.
The comparison between the steady-state cooling limit and the
instantaneous-state cooling limit is also shown in Fig. \ref{Fig2}, which
reveals the advantage of the instantaneous-state cooling limit with a factor
of $\sim\pi{\kappa/(4\left\vert {G}\right\vert )}$.

\section{Cooling dynamics without the rotating-wave approximation}

When the coupling strength $\left\vert {G}\right\vert $ is comparable to the
mechanical resonance frequency ${\omega_{\mathrm{m}}}$, the effect of the
counter-rotating interactions become important. In this case we need to solve
Eq. (\ref{D}), without the the RWA. For convenience, we first consider the
zero-temperature case, where the environmental thermal phonon number
$n_{\mathrm{th}}$ is set to be zero.

\subsection{Zero-temperature case}

For $n_{\mathrm{th}}=0$, all the initial values of the second-order moments in
Eq. (\ref{D}) are zero. With the time evolution, these moments become nonzero
due to the the quantum backaction resulting from the vacuum fluctuations, i.
e., they are created from vacuum by\ the counter-rotating interactions. For
the red sideband resonant case $\Delta^{\prime}=-{\omega_{\mathrm{m}}}$, by
diagonalizing the Hamiltonian Eq. (\ref{HL}), we obtain $H_{L}={\omega}%
_{+}{c_{+}^{\dag}c}_{+}+{\omega}_{-}{c_{-}^{\dag}c}_{-}$, where ${c}_{\pm}$
are the eigenmodes with the corresponding eigenfrequencies ${\omega}_{\pm
}=\sqrt{{\omega_{\mathrm{m}}^{2}}\pm2\left\vert G\right\vert {\omega
_{\mathrm{m}}}}$. Note that the the eigenfrequencies reduce to ${\omega}_{\pm
}={\omega_{\mathrm{m}}}\pm\left\vert G\right\vert $ with the RWA as shown in
the previous section. The rotating-wave interaction is characterized by the
frequency ${\omega}_{+}-{\omega}_{-}\ $[${\sim2}\left\vert G\right\vert $ with
RWA as appeared in Eq. (\ref{NbRWA})], while the counter-rotating-wave
interaction is characterized by the frequency ${\omega}_{+}+{\omega}_{-}$.
Therefore, by taking both interactions into consideration, the system dynamics
is described by two frequencies ${\omega}_{+}\pm{\omega}_{-}$. To provide
quantitative results, we derive the time evolution of the mean phonon number
in this case as \cite{ycliuDC13}%
\begin{equation}
\bar{N}_{b}^{\mathrm{(0)}}(t)\simeq\frac{\left\vert {G}\right\vert ^{2}\left[
1-{e}^{-\frac{{\kappa+\gamma}}{2}t}\cos({\omega}_{+}+{\omega}_{-}%
)t\cos({\omega}_{+}-{\omega}_{-})t\right]  }{{2(\omega_{\mathrm{m}}^{2}%
-4}\left\vert {G}\right\vert ^{2})}. \label{Nb0}%
\end{equation}
Unlike $\bar{N}_{b}^{\mathrm{RWA}}(t)$ as shown in Eq. (\ref{NbRWA}), here
$\bar{N}_{b}^{\mathrm{(0)}}(t)$ does not depend on the environmental thermal
phonon number $n_{\mathrm{th}}$, which reveals that it originates from the
quantum backaction associated with the vacuum fluctuations.

\begin{figure}[t]
\centerline{\includegraphics[width=\columnwidth]{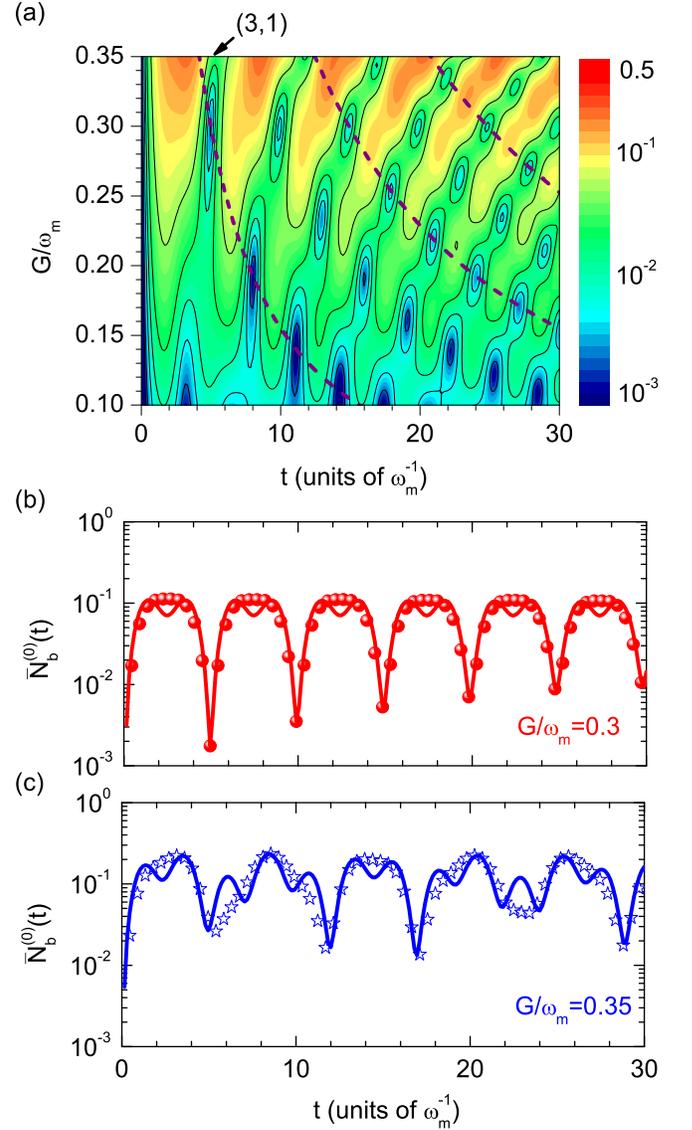}}
\caption{(color online) (a) Time evolution of the mean phonon number $\bar
{N}_{b}^{\mathrm{(0)}}(t)$ as functions of $t$ and $G$. The purple dashed
curves correspond to the cases of $t=\pi/({\omega}_{+}-{\omega}_{-})$,
$3\pi/({\omega}_{+}-{\omega}_{-})$ and $5\pi/({\omega}_{+}-{\omega}_{-})$. The
island in the top left corner is labeled as $(p,q)=(3,1)$. The color bar is in
log scale. The black contour curves denote $\bar{N}_{b}^{\mathrm{(0)}%
}(t)=0.001$, $0.002$, $0.005$, $0.1$, $0.2$ and $0.5$. (b) and (c): $\bar
{N}_{b}^{\mathrm{0}}(t)$ as a function of $t$ for $G{/\omega_{\mathrm{m}}%
}=0.3$ (b) and $0.35$ (c). The circles and stars are numerical results
obtained from Eq. (\ref{D}), and the curves are analytical results calculated
from Eq. (\ref{Nb0}). Other parameters are the same as Fig. \ref{Fig1}(a)
except that $n_{\mathrm{th}}=0$.}%
\label{Fig3}%
\end{figure}

\begin{figure}[t]
\centerline{\includegraphics[width=\columnwidth]{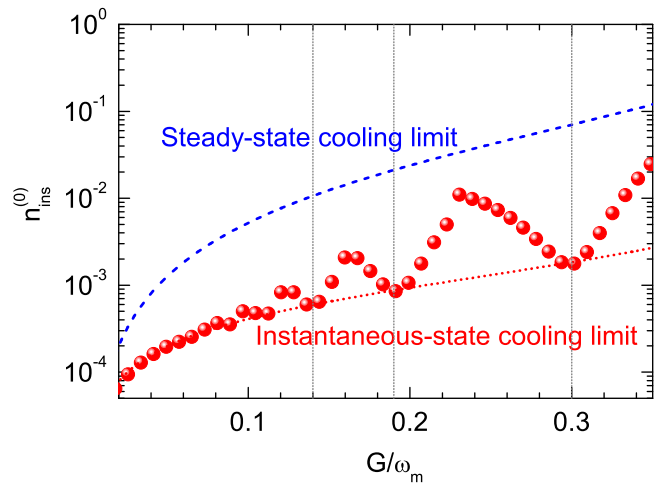}}
\caption{(color online) Instantaneous-state cooling limit $n_{\mathrm{ins}%
}^{\mathrm{(0)}}$ as a function of $G$. The red circles are numerical results
obtained from Eq. (\ref{D}) and the red dotted curve is the analytical result
calculated from Eq. (\ref{Nins0}). The steady-state cooling limit is plotted
for comparison (blue dashed curve). The dotted vertical lines indicate
${G}/{\omega_{\mathrm{m}}=0.14}$, ${0.19}$ and ${0.3}$ from left to right.
Other parameters are the same as Fig. \ref{Fig3}(a).}%
\label{Fig4}%
\end{figure}

In Fig. \ref{Fig3}(a) we present the exact numerical results of $\bar{N}%
_{b}^{\mathrm{(0)}}(t)$ as functions of $t$ and ${G}$ with contour plots. We
note that a number of islands regularly appear in the contour map. This is a
result of the carrier-envelope-type frequency mixing as described by the term
$\cos({\omega}_{+}+{\omega}_{-})t\cos({\omega}_{+}-{\omega}_{-})t$ in Eq.
(\ref{Nb0}), where the carrier frequency ${\omega}_{+}+{\omega}_{-}$
corresponds to the counter-rotating-wave interaction and the envelope
frequency ${\omega}_{+}-{\omega}_{-}$ corresponds to the rotating-wave
interaction. The local minimum value of $\bar{N}_{b}^{\mathrm{(0)}}(t)$ is
obtained when $\cos({\omega}_{+}+{\omega}_{-})t\cos({\omega}_{+}-{\omega}%
_{-})t\simeq1$, which yields the following frequency matching condition
\begin{subequations}
\label{Mat}%
\begin{align}
({\omega}_{+}+{\omega}_{-})t  &  =p\pi,\\
({\omega}_{+}-{\omega}_{-})t  &  =q\pi,
\end{align}
where $p$ and $q$ are both odd integers or both even integers, and $p>q$. For
example, $q=1$, $p=3$, $5$, $7$...; $q=2$, $p=4$, $6$, $8$... We can label
these islands in Fig. \ref{Fig3}(a) as $(p,q)$, where we have marked $(3,1)$
as an example. From Eq. (\ref{Mat}) we derive the corresponding $\left\vert
G\right\vert $ and $t$ for each islands $(p,q)$ as
\end{subequations}
\begin{subequations}
\begin{gather}
\left\vert G\right\vert =\frac{{pq}}{{p}^{2}+q^{2}}{\omega_{\mathrm{m}},}\\
t=\sqrt{{p}^{2}+q^{2}}\frac{\pi}{{2\omega_{\mathrm{m}}}}.
\end{gather}
Because of the dissipation, the optimal minimum value is reached for the
island with the shortest time, which corresponds to $(p,q)=(3,1)$, the labeled
island in the top left corner of Fig. \ref{Fig3}(a). In this case we obtain
$|{G}|/{\omega_{\mathrm{m}}=0.3}$ and $t=\sqrt{{10}}\pi/({2\omega_{\mathrm{m}%
}})$. Some of the other minimum values for $q=1$ [along the leftmost purple
dashed curve in Fig. \ref{Fig3}(a)] can be obtained as $|{G}|/{\omega
_{\mathrm{m}}=p/({p}^{2}+1)=0.19}${, }${0.14}$, $0.11$...($p=5$, $7$, $9$...)

In Figs. \ref{Fig3}(b) and (c) we plot $\bar{N}_{b}^{\mathrm{(0)}}(t)$ for
$G{/\omega_{\mathrm{m}}}=0.3$ and $0.35$, corresponding to the frequency
matched and unmatched regions, respectively. For the former, the time
evolution of phonon number exhibits periodic oscillations, with the minimum
phonon number lower than $10^{-3}$; for the latter, the phonon number does not
show obvious periodicity, and the minimum phonon number is larger than
$10^{-2}$. Although there exist some discrepancies between the analytical
results expressed by Eq. (\ref{Nb0}) and the exact results numerically
computed from Eq. (\ref{D}) in the form of some high frequency fluctuations,
the analytical expression characterizes the minimum phonon number very well,
especially for the frequency-matched case.

For the islands with $q=1$, the minimum phonon number is obtained near the end
of the first half Rabi oscillation cycle, ${t}\simeq\pi/({\omega}_{+}-{\omega
}_{-})$. The instantaneous-state cooling limit in this case is analytically
given by
\end{subequations}
\begin{equation}
n_{\mathrm{ins}}^{\mathrm{(0)}}\simeq\frac{\pi{\kappa}|{G}|}{{8(\omega
_{\mathrm{m}}^{2}-4}|{G}|^{2})},\label{Nins0}%
\end{equation}
where we have neglected the terms containing mechanical dissipation rate
${\gamma}$ since in the experiments typically ${\kappa\gg\gamma}$. We plot
$n_{\mathrm{ins}}^{\mathrm{(0)}}$ as a function of $G$ in Fig. \ref{Fig4}. The
overall trend is that large coupling strength leads to large cooling limit,
which is a result of quantum backaction heating in the strong coupling regime.
The analytical expression Eq. (\ref{Nins0}) describes this trend quite well.
Meanwhile, the oscillations are signatures of frequency matching, where
$\left\vert {G}\right\vert /{\omega_{\mathrm{m}}=0.3}$, ${0.19}${\ and
}${0.14}$ correspond to the frequency-matched regions with low phonon number.
The steady-state cooling limit reads $\left\vert {G}\right\vert ^{2}%
/[{2(\omega_{\mathrm{m}}^{2}-4}\left\vert {G}\right\vert ^{2})]$, which is
also plotted in Fig. \ref{Fig4} for comparison.

\subsection{Finite-temperature case}

\begin{figure}[t]
\centerline{\includegraphics[width=\columnwidth]{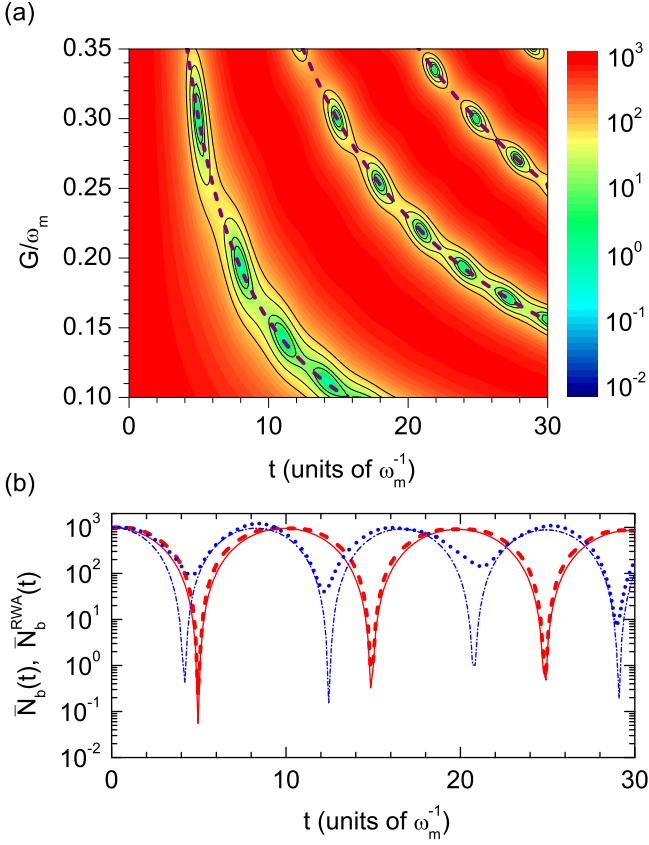}}
\caption{(color online) (a) Time evolution of the mean phonon number $\bar
{N}_{b}(t)$ as functions of $t$ and $G$. The purple dashed curves correspond
to the cases of $t=\pi/({\omega}_{+}-{\omega}_{-})$, $3\pi/({\omega}%
_{+}-{\omega}_{-})$ and $5\pi/({\omega}_{+}-{\omega}_{-})$. The color bar is
in log scale. The black contour curves denote $\bar{N}_{b}(t)=10$, $20$, $50$
and $100$. (b) $\bar{N}_{b}(t)$ as a function of $t$ for $G{/\omega
_{\mathrm{m}}}=0.3$ (red dashed curve) and $0.35$ (blue dotted curve), which
are numerical results obtained from Eq. (\ref{D}); $\bar{N}_{b}^{\mathrm{RWA}%
}(t)$ as a function of $t$ for $G{/\omega_{\mathrm{m}}}=0.3$ (red solid curve)
and $0.35$ (blue dash-dotted curve), which are numerical results obtained from
Eq. (\ref{DRWA}). Other parameters are the same as Fig. \ref{Fig1}(a).}%
\label{Fig5}%
\end{figure}

For finite-temperature case $n_{\mathrm{th}}\neq0$, the phonon number
originates from both the modified mechanical dissipation and quantum
backaction. By solving Eq. (\ref{D}), in Fig. \ref{Fig5}(a) we plot the exact
numerical results of the time evolution of the mean phonon number $\bar{N}%
_{b}(t)$ as functions of $t$ and $G$. The contour map shows that $\bar{N}%
_{b}(t)$ also exhibits\emph{ }Rabi-like oscillations and a number of islands.
The main features of the Rabi-like oscillation is similar to the RWA case as
shown in Fig. \ref{Fig1}, while the island appears only when both $p$ and $q$
are odd integers, corresponding to the Rabi-like oscillation dips. This is
because the contribution of $\bar{N}_{b}^{\mathrm{(0)}}(t)$ is hidden by
$\bar{N}_{b}^{\mathrm{RWA}}(t)$ except for the regions where $\bar{N}%
_{b}^{\mathrm{RWA}}(t)$ reaches the minimum, i. e., $\cos({\omega}_{+}%
-{\omega}_{-})t\sim-1$, corresponding to odd $q$. In Fig. \ref{Fig5}(b) we
compare $\bar{N}_{b}(t)$ and $\bar{N}_{b}^{\mathrm{RWA}}(t)$ for both
$G{/\omega_{\mathrm{m}}}=0.3$ and $0.35$, which demonstrates the significance
of frequency matching. For $G{/\omega_{\mathrm{m}}}=0.3$ (frequency-matched
case), the difference between $\bar{N}_{b}(t)$ and $\bar{N}_{b}^{\mathrm{RWA}%
}(t)$ is small and the minimum phonon number is less than $10^{-1}$. However,
for $G{/\omega_{\mathrm{m}}}=0.35$ (frequency-unmatched case), $\bar{N}%
_{b}(t)$ is different from $\bar{N}_{b}^{\mathrm{RWA}}(t)$ near the
Rabi-oscillation dips, since the frequency matching condition is not
satisfied. In this case the quantum backaction plays an important role and the
minimum phonon number is as large as $10^{2}$. In Fig. \ref{Fig6}(a), we
present the instantaneous-state cooling limit $n_{\mathrm{ins}}$ [obtained
near ${t}=\pi/({\omega}_{+}-{\omega}_{-})$] as a function of $G$ (blue stars),
which reveals obvious difference between the frequency matched\ and unmatched cases.

\section{Reduced instantaneous-state cooling limits}

\begin{figure}[tb]
\centerline{\includegraphics[width=\columnwidth]{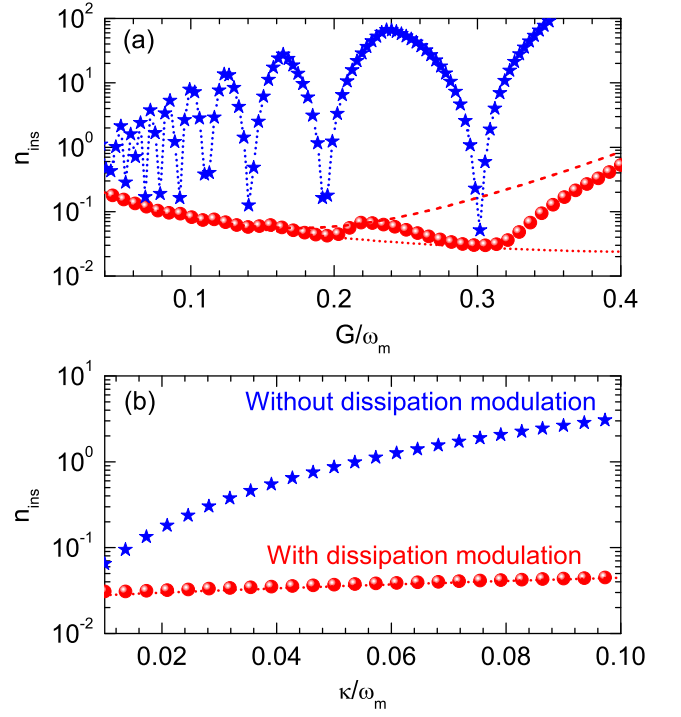}}
\caption{(color online) (a) Instantaneous-state cooling limits
$n_{\mathrm{ins}}$ as a function of $G$ for ${\kappa/{\omega_{\mathrm{m}}%
}=0.01}$. (b) $n_{\mathrm{ins}}$ as a function of ${\kappa}$ for
${G/{\omega_{\mathrm{m}}}=0.3}$. The red circles are numerical results
obtained from Eq. (\ref{D}) with dynamic cavity dissipation modulation; the
red dashed and dotted curves are the analytical results calculated from Eqs.
(\ref{nins}) and (\ref{ninsopt}); the blue stars are the numerical results
obtained from Eq. (\ref{D}) without dynamic cavity dissipation modulation.
Other parameters are the same as Fig. \ref{Fig1}(a).}%
\label{Fig6}%
\end{figure}

By employing the dynamic dissipative cooling approach \cite{ycliuDC13}, the
instantaneous-state cooling limit can be can be significantly reduced. In Fig.
\ref{Fig6}(a) we plot the reduced instantaneous-state cooling limits with
dynamic cavity dissipation modulation as a function of $G$ for ${\kappa
/\omega_{\mathrm{m}}=0.01}$ (red circles). Analytical derivations lead to the
cooling limits \cite{ycliuDC13}
\begin{subequations}
\begin{align}
n_{\mathrm{ins}}  &  \simeq\frac{\pi{\gamma}n_{\mathrm{th}}}{4\left\vert
{G}\right\vert }+\frac{\pi^{2}\left\vert {G}\right\vert ^{4}}{{(\omega
_{\mathrm{m}}^{2}-}\left\vert {G}\right\vert ^{2}){(\omega_{\mathrm{m}}^{2}%
-4}\left\vert {G}\right\vert ^{2})},\label{nins}\\
n_{\mathrm{ins}}^{\mathrm{opt}}  &  \simeq\frac{\pi{\gamma}n_{\mathrm{th}}%
}{4\left\vert {G}\right\vert }+\frac{\pi{\kappa}|{G}|}{{8(\omega_{\mathrm{m}%
}^{2}-4}|{G}|^{2})}, \label{ninsopt}%
\end{align}
where $n_{\mathrm{ins}}^{\mathrm{opt}}$ denotes the instantaneous-state
cooling limits for the optimal coupling strength $G$ (frequency-matched case),
and\ $n_{\mathrm{ins}}$ corresponds to the frequency-unmatched case. It
reveals that these two expressions set the upper and lower bounds of the
instantaneous-state cooling limits, as shown in Fig. \ref{Fig6}(a). It is
worth noting that the lower bound is the sum of the two cooling limits given
by Eq. (\ref{NinsRWA}) and Eq. (\ref{Nins0}), i. e., $n_{\mathrm{ins}%
}^{\mathrm{opt}}=n_{\mathrm{ins}}^{\mathrm{RWA}}+n_{\mathrm{ins}%
}^{\mathrm{(0)}}$. Typically, the minimum cooling limit is obtained for
$\left\vert {G}\right\vert /{\omega_{\mathrm{m}}}=0.3$, as shown in Fig.
\ref{Fig6}(a). In Fig. \ref{Fig6}(b) the instantaneous-state cooling limits as
functions of the cavity decay rate ${\kappa}$ for the optimal coupling
strength $\left\vert {G}\right\vert /{\omega_{\mathrm{m}}=0.3}$ are plotted.
It shows that a small $\kappa$ is preferred to obtain a low cooling limit.
Therefore, to attain the best cooling performance in the strong coupling
regime, the frequency matching condition should be satisfied and a small
cavity decay rate is required. This is in contrast to the steady-state cooling
limit which requires an optimal value for ${\kappa}$ to balance the quantum
limit and cavity bandwidth limitation \cite{SCPRL08,SCNJP08}.

\section{Conclusions}

In summary, we have examined the backaction cooling of mesoscopic mechanical
resonators in the strong coupling regime. When the rotating-wave approximation
is satisfied, the mean phonon number undergoes simple damped Rabi-like
oscillations with the Rabi frequency of $2\left\vert {G}\right\vert $ and the
exponential damping envelope scales as ${e}^{-\frac{{\kappa+\gamma}}{2}t}$. At
the first half oscillation cycle, the minimum mean phonon number is derived as
$\pi{\gamma}n_{\mathrm{th}}/(4\left\vert {G}\right\vert )$. For large coupling
strength where the rotating-wave approximation fails, the mean phonon number
oscillates as functions of both the evolution time and the coupling strength,
which is a result of frequency mixing. Under the frequency matching condition,
the optimal coupling strengths are derived as $\left\vert {G}\right\vert
/{\omega_{\mathrm{m}}}\simeq0.3$, $0.19$, $0.14$... By employing the dynamic
dissipative approach \cite{ycliuDC13}, the reduced instantaneous-state cooling
limits reach the lower bound $\pi{\gamma}n_{\mathrm{th}}/(4\left\vert
{G}\right\vert )+\pi{\kappa}|{G}|/[{8(\omega_{\mathrm{m}}^{2}-4}|{G}|^{2})]$.
This provides a guideline for achieving the lowest cooling limit, which is
reached when the frequency matching condition is satisfied and when the cavity
decay rate ${\kappa}$ is small. Compared with the steady-state cooling limit,
the unique advantage is that it does not require an optimal value for
${\kappa}$, allowing for long-coherence-time quantum operations deeply in the
strong coupling regime. The parameter ranges $\gamma\ll\kappa<G<\omega
_{\mathrm{m}}/2$ can be realized in various optomechanical systems, for
example, in the microtoroid system studied in Ref. \cite{SCNat12}, with the
parameters $\omega_{\mathrm{m}}/2\pi\sim78$ $\mathrm{MHz}$, $\kappa/2\pi
\sim7.1$ $\mathrm{MHz}$, $\gamma/2\pi\sim10$ $\mathrm{kHz}$ and $G/2\pi
\sim11.4$ $\mathrm{MHz}$, or in the superconducting aluminium membrane system
studied in Ref. \cite{SCNat13}, with the parameters $\omega_{\mathrm{m}}%
/2\pi\sim10.5$ $\mathrm{MHz}$, $\kappa/2\pi\sim320$ $\mathrm{kHz}$,
$\gamma/2\pi\sim35$ $\mathrm{Hz}$ and $G>\kappa$.
\end{subequations}
\begin{acknowledgments}
This work is supported by the 973 program (2013CB328704, 2013CB921904), NSFC
(11004003, 11222440, and 11121091), and RFDPH (20120001110068). Y.C.L is
supported by the Scholarship Award for Excellent Doctoral Students granted by
the Ministry of Education. Y.F.S was supported by the National Fund for
Fostering Talents of Basic Science (Grants No. J1030310 and No. J1103205)
\end{acknowledgments}

\appendix

\section{}

In Eq. (\ref{D}), the matrices $\mathbf{M}$ and $\mathbf{N}$ are given by
\begin{widetext}
\begin{align}
\mathbf{M}  & \mathbf{=}\left(
\begin{array}
[c]{ccccc}%
-{\kappa} & 0 & -iG & iG^{\ast} & iG^{\ast}\\
0 & -{\gamma} & iG & -iG^{\ast} & iG^{\ast}\\
-iG^{\ast} & iG^{\ast} & -i\left(  \Delta^{\prime}+{\omega_{\mathrm{m}}%
}\right)  -\frac{{\kappa+\gamma}}{2} & 0 & 0\\
iG & -iG & 0 & i\left(  \Delta^{\prime}+{\omega_{\mathrm{m}}}\right)
-\frac{{\kappa+\gamma}}{2} & 0\\
-iG & -iG & 0 & 0 & i\left(  \Delta^{\prime}-{\omega_{\mathrm{m}}}\right)
-\frac{{\kappa+\gamma}}{2}\\
iG^{\ast} & iG^{\ast} & 0 & 0 & 0\\
0 & 0 & 0 & -2iG & -2iG\\
0 & 0 & 2iG^{\ast} & 0 & 0\\
0 & 0 & -2iG & 0 & -2iG^{\ast}\\
0 & 0 & 0 & 2iG^{\ast} & 0
\end{array}
\right.  \nonumber\\
& \left.
\begin{array}
[c]{ccccc}%
-iG & 0 & 0 & 0 & 0\\
-iG & 0 & 0 & 0 & 0\\
0 & 0 & -iG & iG^{\ast} & 0\\
0 & iG^{\ast} & 0 & 0 & -iG\\
0 & -iG^{\ast} & 0 & -iG & 0\\
-i\left(  \Delta^{\prime}-{\omega_{\mathrm{m}}}\right)  -\frac{{\kappa+\gamma
}}{2} & 0 & iG & 0 & iG^{\ast}\\
0 & 2i\Delta^{\prime}-{\kappa} & 0 & 0 & 0\\
2iG^{\ast} & 0 & -2i\Delta^{\prime}-{\kappa} & 0 & 0\\
0 & 0 & 0 & -2i{\omega_{\mathrm{m}}}-{\gamma} & 0\\
2iG & 0 & 0 & 0 & 2i{\omega_{\mathrm{m}}}-{\gamma}%
\end{array}
\right),
\end{align}
\end{widetext}
\begin{equation}
\mathbf{N}\mathbf{=(}0,{\gamma}n_{\mathrm{th}},0,0,-iG,iG^{\ast},0,0,0,0)^{T}.
\end{equation}
Initially, the mean phonon number is equal to the bath thermal phonon number
and other second-order moments are zero, i. e., $\mathbf{V}%
(t=0)=(0,n_{\mathrm{th}},0,0,0,0,0,0,0,0)^{T}$.

\end{document}